\documentstyle[sprocl]{article}

\def\be{\begin{equation}}
\def\ee{\end{equation}}
\def\bea{\begin{eqnarray}}
\def\eea{\end{eqnarray}}
\begin{document}

\title{QUARK FRAGMENTATION AND OFF-DIAGONAL\\
 HELICITY DENSITY MATRIX ELEMENTS\\
 FOR VECTOR MESON PRODUCTION}

\author{M. Anselmino$^a$, M. Bertini$^a$, F. Murgia$^b$,
        P. Quintairos$^{c}$}

\address{
\baselineskip=7pt
$^a$Dipartimento di Fisica Teorica, Universit\`a di Torino and \\
 INFN, Sezione di Torino, Via P. Giuria 1, 10125 Torino, Italy\\
$^b$Dipartimento di Fisica, Universit\`a di Cagliari and \\
INFN, Sezione di Cagliari, C.P. n. 170, I-09042 Monserrato (CA), Italy\\
$^c$Centro Brasileiro de Pesquisas F\'{\i}sicas \\
R. Dr. Xavier Sigaud 150, 22290-180, Rio de Janeiro, Brazil}

%%%%%%%%%%%%%%%%%%%%%%%%%%%%%%%%%%%%%%%%%%%%%%%%%%%%%%%%%%%%%%
% You may repeat \author \address as often as necessary      %
%%%%%%%%%%%%%%%%%%%%%%%%%%%%%%%%%%%%%%%%%%%%%%%%%%%%%%%%%%%%%%

\vspace{-3pt}

\maketitle\abstracts{
As confirmed by some recent LEP data on $\phi$, $K^*$ and $D^*$ production,
final state interactions in quark fragmentation may give origin to
non-zero values of the off-diagonal element $\rho_{1,-1}$ of the helicity 
density matrix of vector mesons produced in $e^+ e^-$ annihilations: 
we give estimates for $\rho_{1,-1}$ of vector mesons  with a large $x_E$ and 
collinear with the parent jet, relating its size and sign to the associated 
hard constituent dynamics. We mention possible non-zero values of 
$\rho_{1,-1}$ in several other processes.}

\vspace{-5pt}

Due to final state interactions and a coherent $q\bar{q}$ fragmentation
process, the off-diagonal matrix element $\rho_{1,-1}$ of vector
mesons inclusively produced in $e^+e^-$ annihilation may be sizeably
different from zero \cite{akp,aamr}.
On the contrary, the commonly adopted incoherent fragmentation scheme
of a single, independent quark leads to zero values for such off-diagonal
elements. These coherent fragmentation effects have recently received some
confirmation from experimental results at LEP \cite{opal}.
Numerical estimates of $\rho_{1,-1}(V)$ in the coherent fragmentation of 
$q\bar{q}$ pairs in $e^+e^-\to q\bar{q} \to V\,X$ processes have been
given \cite{abmq} for several vector mesons produced in two jet events, 
provided they have a large energy fraction $x_E=E_V/E_{beam}$,
and a small transverse momentum $p_T$ inside the jet.

The polarization state of $V$ is described by its helicity density
matrix $\rho(V)$, whose general expression is:

\vspace{-8pt}
\begin{equation}
\rho_{\lambda^{\,}_V\lambda'_V}(V) = \frac{1}{N_V}\sum_{q,X,\lambda's}
{\cal D}_{\lambda^{\,}_V\lambda^{\,}_X;\lambda^{\,}_q\lambda^{\,}_{\bar{q}}} \>
{\cal D}^*_{\lambda'_V\lambda^{\,}_X;\lambda'_q\lambda'_{\bar{q}}} \>
\rho_{\lambda^{\,}_q\lambda^{\,}_{\bar{q}};\lambda'_q\lambda'_{\bar{q}}}
(q\bar{q}) \label{rv}
\end{equation}

\vspace{-8pt}
\noindent where $N_V$ is such that $\mbox{Tr}(\rho)=1$,
and the ${\cal D}$'s are unknown, non perturbative helicity amplitudes
for the process, $q\bar{q}\to V\,X$.
The $q\bar{q}$ density matrix is in turn given by

\vspace{-8pt}
\begin{equation}
\rho_{\lambda^{\,}_q\lambda^{\,}_{\bar{q}};
\lambda'_q\lambda'_{\bar{q}}}(q\bar{q})=
\frac{1}{N_q}\sum_{\lambda_{e^-}\lambda_{e^+}}
M_{\lambda^{\,}_q\lambda^{\,}_{\bar{q}};\lambda_{e^-}\lambda_{e^+}}
M^*_{\lambda'_q\lambda'_{\bar{q}};\lambda_{e^-}\lambda_{e^+}}
\label{rq}
\end{equation}
\vspace{-8pt}

\noindent where the $M$'s are the helicity amplitudes for the process
$e^-e^+\to q\bar{q}$.
It is easy to show that in case of a single, independent, collinear quark
fragmentation all off-diagonal elements of $\rho_{\lambda^{\,}_V\lambda'_V}(V)$ 
vanish.

Although the amplitudes
${\cal D}_{\lambda_V\lambda_X;\lambda_q\lambda_{\bar{q}}}$ 
are unknown, it is possible \cite{abmq}, at least 
in some kinematical regimes, to express
$\rho_{\lambda^{\,}_V\lambda'_V}(V)$ by means of
essentially only one unknown, non perturbative, quantity.
In details, the procedure is the following: {\it i)} Consider only
vector mesons collinear with the parent jet ($p_T/(x_E\sqrt{s})\to 0$).
This implies that the diagonal matrix elements are the same (up to
corrections of the order $[p_T/(x_E\sqrt{s})^2$] as in the incoherent
fragmentation picture; moreover, the only non-vanishing, off-diagonal
matrix elements are, to the same order, $\rho_{\pm 1,\mp 1}(V)$.
The surviving combinations of ${\cal D}$ amplitudes in
Eq.~(\ref{rv}) are:
$\sum_X|{\cal D}_{1,0;+,-}|^2 = D_{q,+}^{V,+1}$;
$\sum_X|{\cal D}_{0,-1;+,-}|^2 = D_{q,+}^{V,0}$;
$\sum_X|{\cal D}_{-1,-2;+,-}|^2 = D_{q,+}^{V,-1}$;
$\sum_X{\cal D}_{1,0;+,-}{\cal D}^*_{-1,0;-,+}
\simeq D_{q,+}^{V,+1}$, where parity conservation and dominance
of the $S_X=0$ contribution have been used.
{\it ii)} By limiting ourself to the region $x_E \ge 0.5$,
we may reasonably assume that the $q\bar{q}$ pair fragments into
a meson $V$ only if $q$ is a valence quark for $V$ itself.
In first approximation, we can also set
$D_{q,+}^{V,-1}=0$, $D_{q,+}^{V,0} \simeq \alpha_q^V D_{q,+}^{V,+1}$,
as for $SU(6)$ vector meson wavefunctions with no orbital
angular momentum.
{\it iii)} Finally, we assume that $\alpha_q^V \simeq \alpha^V$
independently of the valence quark flavour.
This condition is well satisfied
for vector mesons with valence quarks all of the same flavour
({\it e.g.}, for $\phi$ meson),
for vector mesons where one valence quark dominates ({\it e.g.}
$D^{*+}$, $B^{*+}$), or with only $u$, $\bar u$,$d$, $\bar d$,valence
quarks ({\it e.g.} $\rho$). It may be weaker for mesons like
$K^{*+}$; in this case the results can be similar to those for
$\rho^+$ (if $u$ and $\bar s$ contribute almost equally) or to
those for $B^{*+}$ (if $\bar s$ contribution dominates).

These approximations lead to very simple results for $\rho(V)$:

\vspace{-8pt}
\begin{equation}
\rho_{00}(V)\simeq\frac{\alpha^V}{1+\alpha^V} \quad \Longrightarrow \quad 
\alpha^V\simeq\frac{\rho_{00}(V)}{1-\rho_{00}(V)}
\label{r00}
\end{equation}
\vspace{-8pt}
 
\vspace{-8pt}
\begin{equation}
\rho_{1,-1}(V)\simeq\Bigl[1-\rho_{00}(V)\Bigr] \frac {1}{n_{_V}}\sum_q
\rho_{+-;-+}(q\bar{q})
\label{r11}
\end{equation}
where $n_{_V}$ is the number of leading quarks contributing.

The unknown parameter $\alpha^V$ can thus be estimated from
experimental data on $\rho_{00}(V)$. Notice that, in $SU(6)$,
$\alpha^V=1/2$, corresponding to $\rho_{00}(V)=1/3$, {\it i.e.} no
spin alignment, $A=(3\rho_{00}-1)/2 = 0$.

Both $\rho_{00}$ and $\rho_{1,-1}$ can be measured via the angular 
distribution of the vector meson decays into two pseudoscalar particles. 
At LEP energies ($\sqrt{s}=M_Z$), computing $\rho_{+-;-+}(q\bar{q})$
as given by $q\bar q$ annihilation into $Z_0$, one gets \cite{abmq}

\vspace{-8pt}
\begin{equation}
\rho_{1,-1}(V)\simeq K^Z_V\Bigl[1-\rho_{00}(V)\Bigr]
\frac{\sin^2\theta}{1+\cos^2\theta}
\label{rz}
\end{equation}
\vspace{-8pt}

\noindent with $K^Z_{\rho,K^{*\pm}}=-0.265$, $K^Z_{K^{*0},\phi,B^*}=-0.170$,
$K^Z_{D^*}=-0.360$. We want to stress that
at low energies, where the e.m. contribution dominates, one gets
$K^\gamma_V=1/2$ for all mesons.

The following table compares our results with recent experimental data from
LEP \cite{opal} (after averaging the $\rho_{1,-1}(V)$
values over the production angle of the vector meson).

\vspace{6pt}

\begin{center}
 \begin{tabular}{ccc}
  \hline
  \multicolumn{2}{c}{Exp (OPAL Collab. \cite{opal})} & Theory \\
  \hline
  $\phi  \; (x_E > 0.7)$     &  $-0.11 \pm 0.07$   &  $-$0.042 \\
  $D^*   \; (x_E > 0.5)$     &  $-0.039 \pm 0.016$ &  $-$0.116 \\
  $K^{*0}\;(0.5< x_E < 0.7)$ &  $-0.05\pm 0.04$    &  $-$0.043  \\
  $K^{*0}\;(0.7< x_E < 1.0)$ &  $-0.08\pm 0.07$    &  $-$0.031  \\
  \hline
 \end{tabular}
\end{center}

\vspace{6pt}

Our results compare well with data from OPAL Collaboration. We notice
however that DELPHI Collaboration \cite{delp} measured $\rho_{1,-1}$ for
$\rho$, $K^{*0}$, $\phi$ mesons over different $x_E$ ranges, finding
values compatible with zero, although error bars are comparable with
our estimated theoretical values. We clearly need more experimental,
dedicated investigations, possibly with an accurate selection of
the kinematical range where our predictions are supposed to be most
reliable. If the OPAL results will be confirmed, this should be a
strong indication for the relevance of coherent fragmentation processes.

Let us finally stress that
our model and its simple prediction, Eq.~(\ref{rz}), can be further
tested in present or near-future experiments. For example, in
$e^+e^-$ annihilation at low energies, we predict $K^\gamma_V=1/2$,
that is $\rho_{1,-1}$ positive for all vector mesons.
A detailed account for inclusive vector meson production in
$NN$, $\gamma N$, $\ell N$ collisions has also been given \cite{abmp},
while the case of $\gamma\gamma$ collisions is under study \cite{abm2}.

\end{document}